\documentclass[aps,prd,twocolumn,floatfix,preprintnumbers,nofootinbib,superscriptaddress]{revtex4-1}
\usepackage[margin=0.7in]{geometry}
\usepackage[utf8]{inputenc}

\usepackage{physics}
\usepackage{amsmath,amssymb,amsthm,amsfonts}
\usepackage[final]{graphicx}
\usepackage{subcaption}
\usepackage{lipsum}
\usepackage{mathtools}
\usepackage{enumitem,tabulary}
\usepackage{indentfirst}
\usepackage[english]{babel}
\usepackage{textcomp}
\usepackage{multirow}
\usepackage{tikz}
\usepackage{tipa}
\usepackage{CJKutf8}
\usepackage{feynmf}
\usepackage{slashed}
\usepackage{braket}
\usepackage[toc,page]{appendix}
\usepackage{url}
\usepackage{natbib}
\usepackage{graphicx}
\usepackage{mathrsfs}  
\usepackage{cancel}
\usepackage[normalem]{ulem}
\usepackage{array}
\usepackage{booktabs}
\usepackage{verbatim}
\usepackage{ragged2e}

\usetikzlibrary{positioning,decorations.pathmorphing,decorations.markings,arrows}
\usepackage{latexsym,amsfonts,color,amsthm}
\usepackage[compat=1.1.0]{tikz-feynman}
\RequirePackage[colorlinks=true
,urlcolor=blue
,anchorcolor=blue
,citecolor=blue
,filecolor=blue
,linkcolor=blue
,menucolor=blue
,linktocpage=true
,pdfproducer=medialab
,pdfa=true
]{hyperref}

\begin{document}
\title{New Physics at Neutron Beam Dump}

\author{P. S. Bhupal Dev}
\email{bdev@wustl.edu}
\affiliation{Department of Physics and McDonnell Center for the Space Sciences, Washington University, St.~Louis, MO 63130, USA}

\author{Bhaskar Dutta}
\email{dutta@tamu.edu}
\affiliation{Mitchell Institute for Fundamental Physics and Astronomy, Department of Physics and Astronomy, Texas A\&M University, College Station, TX 77843, USA}

\author{Tao Han}
\email{than@pitt.edu}
\affiliation{PITT PACC, Department of Physics and Astronomy, University of Pittsburgh, Pittsburgh, PA 15260, USA}

\author{Aparajitha Karthikeyan}
\email{aparajitha\_96@tamu.edu}
\affiliation{Mitchell Institute for Fundamental Physics and Astronomy, Department of Physics and Astronomy, Texas A\&M University, College Station, TX 77843, USA}

\author{Doojin Kim}
\email{doojin.kim@tamu.edu}
\affiliation{Mitchell Institute for Fundamental Physics and Astronomy, Department of Physics and Astronomy, Texas A\&M University, College Station, TX 77843, USA}

\author{Hyunyong Kim}
\email{hyunyong@cern.ch}
\affiliation{Mitchell Institute for Fundamental Physics and Astronomy, Department of Physics and Astronomy, Texas A\&M University, College Station, TX 77843, USA}

\begin{abstract}
     We find a new utility of neutrons, usually treated as an experimental nuisance causing unwanted background, in probing new physics signals. The new physics signals can either be radiated from neutrons  (neutron bremsstrahlung) or appear through secondary particles from neutron-on-target interactions, dubbed ``neutron beam dump''. As a concrete example, we take the FASER/FASER2 experiment as a ``factory'' of high-energy neutrons that interact with the iron dump. We find that if new physics particles had the same interaction strength to protons and neutrons, their production rate via neutron-initiated bremsstrahlung would be comparable to that via proton-initiated ones, in terms of the resulting flux and the range of couplings that can be probed. The neutron bremsstrahlung can be used, for instance, to probe dark gauge bosons with non-zero neutron coupling. In particular, we investigate protophobic gauge bosons and find that FASER/FASER2 can probe new parameter space. We also illustrate the possibility of neutron-induced secondary particles by considering axion-like particles with electron couplings. We conclude that the physics potential of FASER/FASER2 in terms of new physics searches can be greatly extended and improved with the inclusion of neutron interactions.
     
\end{abstract}

\maketitle

\section{Introduction} 
Light (sub-GeV) new physics particles feebly interacting with Standard Model (SM) particles are receiving increasing attention~\cite{Agrawal:2021dbo} as they can address various phenomenological issues including experimental anomalies, e.g.,~LSND~\cite{LSND:2001aii}, MiniBooNE~\cite{MiniBooNE:2008yuf, MiniBooNE:2018esg, MiniBooNE:2020pnu}, muon $g-2$~\cite{Muong-2:2006rrc, Muong-2:2021ojo, 
Muong-2:2023cdq}, and ATOMKI~\cite{Krasznahorkay:2015iga, Krasznahorkay:2019lyl, Krasznahorkay:2021joi, Krasznahorkay:2022pxs, Krasznahorkay:2023sax} anomalies.
In particular, portal scenarios in connection to low-mass dark matter strongly motivate sub-GeV-range bosonic mediators of this sort~\cite{Batell:2022xau}. 
Due to their feebly-interacting nature and low-mass scale, accelerator-based experiments utilizing high-intensity particle beams, i.e., beam-dump-type or fixed-target experiments, are better suited than high-energy colliders in their searches. 
These intensity-frontier experiments conventionally use incident beam particles (e.g., protons and electrons) and neutral mesons (e.g., $\pi^0$, $\eta$, $\eta'$) as the sources for producing the above-mentioned light dark-sector mediators. 
Recent studies pointed out the use of charged mesons (e.g.~$\pi^\pm$, $K^\pm$)~\cite{Pang:1973rxr, Carlson:2012pc, Dutta:2021cip, Dev:2023rqb, Dutta:2023fnl} and secondary photons and electrons~\cite{Capozzi:2023ffu, CCM:2021jmk} as well. 
The large beam intensity essentially allows for copious production of such source particles, hence the mediators can be produced abundantly.

In this paper, we point out that {\it neutrons} can serve as a new 
promising route, mostly overlooked thus far, to search for new physics signatures in beam-dump experiments. It has been standard lore that neutrons are a nuisance, as they induce unwanted backgrounds, 
especially in short-baseline experiments. However, putting aside the fact that neutrons are electromagnetically neutral, they are as rich in hadronic properties as protons. Therefore, they are (almost) equally capable of inducing new physics as well as SM particles; this new production channel allows us to explore a new set of models, in particular protophobic ones, at dump facilities and provide information complementary to existing searches. 

A variety of experiments can make use of neutrons to search for particles that exist beyond the Standard Model (BSM). Examples include forward physics facilities (FPFs)~\cite{Feng:2022inv} such as FASER~\cite{FASER:2018bac}, FLArE~\cite{Batell:2021blf}, SND@LHC~\cite{SHiP:2020sos}, FORMOSA~\cite{Foroughi-Abari:2020qar}, and stopped-pion neutrino experiments like KARMEN~\cite{KARMEN:1991vkr, Kleinfeller:1994xf}, CCM~\cite{Aguilar-Arevalo:2023kvr, CCM:2021jmk}, COHERENT~\cite{Akimov:2022oyb}, and JSNS$^2$~\cite{Ajimura:2017fld}.
While experiments using lower-energy particle beams would provide interesting opportunities, here we focus on the FASER/FASER2 and FLArE setup as they are characterized by distinctive features.

\begin{figure*}[t]
    \centering
    \includegraphics[width = 0.9\textwidth]{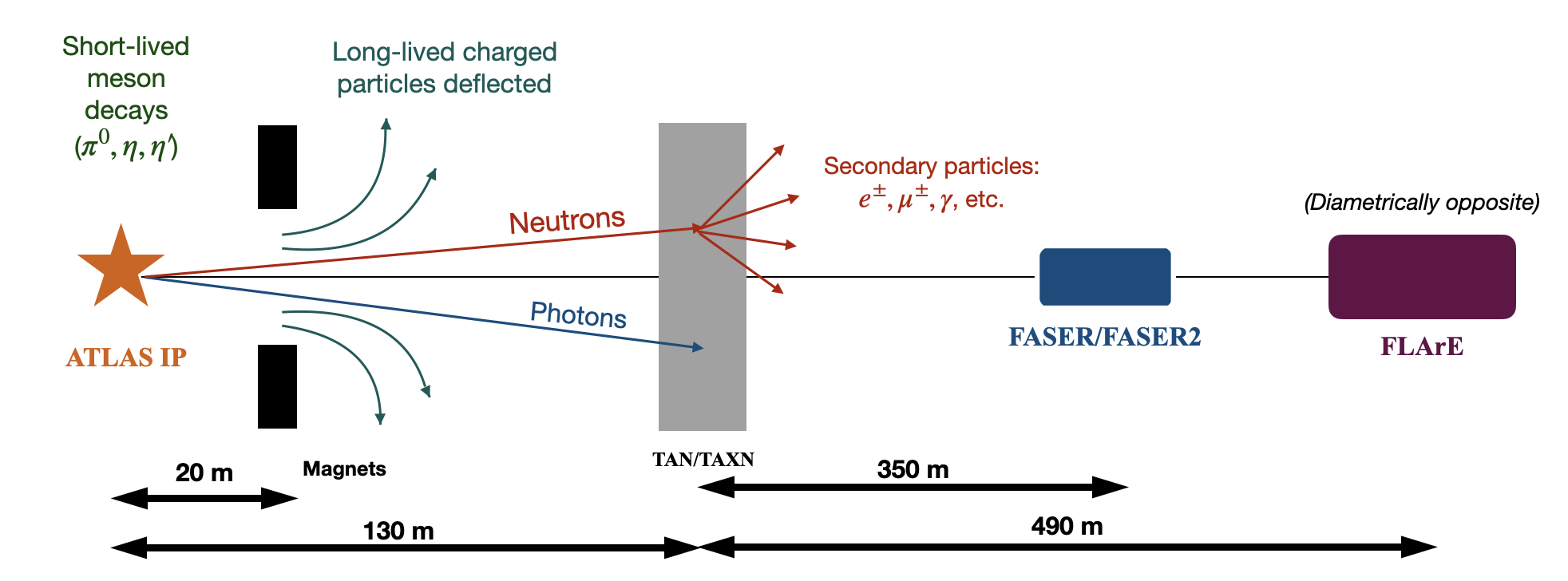}
    \captionsetup{justification=Justified, singlelinecheck=false}
    \caption{A schematic depiction of our idea implemented at the FASER/FASER2, and FLArE (which is to be placed on the diametrically opposite side of FASER/FASER2) experiment (not to scale).}  
    \label{fig:fasersketch}
\end{figure*}

The ongoing FASER experiment and the upcoming FASER2 and FLArE experiments are a part of the FPF program, which aims to explore BSM physics at high energies in a unique way~\cite{Feng:2022inv}. They seek signals coming from the interaction point (IP) at ATLAS on the LHC beam line. Table~\ref{tab:FASER} summarizes the detector details as well as luminosities at which they are operating/set to operate.
Figure~\ref{fig:fasersketch} (not to scale) schematically depicts the experimental setup of the above-mentioned detectors as well as possible BSM sources. Existing studies which discuss searches for new light scalars~\cite{Feng:2017vli, Kelly:2021mcd, Kling:2022uzy}, heavy neutral leptons~\cite{Kling:2018wct}, axion-like particles (ALPs)~\cite{Kling:2022ehv,Feng:2018pew}, and dark gauge bosons~\cite{FASER:2018eoc, Kling:2020iar, Araki:2020wkq, Batell:2021snh, FASER:2023tle} employ (short-lived) mesons, which decay before the LHC magnets, and photons, which survive beyond them. We note that a large number of neutrons produced at the IP are also undeflected by the magnets due to their electrically neutral nature, and are dumped to TAN/TAXN, along with photons.
Originating from the LHC proton-proton collisions, such neutrons are as energetic as $\mathcal{O}$(TeV), and inside TAN/TAXN they can result in secondary particles carrying large enough energy to produce BSM particles. 
This scheme introduces a new experimental concept, {\it neutron beam-dump experiments}.
In this context, we propose extending the BSM search program of FASER by including this neutron channel. 

\begin{table}[t!]
    \begin{tabular}{|c|c|c|c|c|}
        \hline 
        Detector & Radius [m] & Depth [m]  & Luminosity [fb$^{-1}$]  & Dump \\
        \hline 
        FASER & $0.1$ & $1.5$ & $150$ &  TAN  \\
        \hline
        FASER 2 & $1$ & $5$ & $3000$ &  TAXN  \\
        \hline
        FLArE & $0.9$ & $30$ & $3000$ &  TAXN  \\
        \hline
    \end{tabular}
    \captionsetup{justification=Justified}
    \caption{Detector specifications (radius and depth) and luminosity of the experiments utilized in this study. The depth here is the length of the allowed decay volume of each detector.}
    \label{tab:FASER}
\end{table}

The paper is organized as follows. In Sec.~\ref{sec:nbrem}, we will sketch the important details of calculating the cross-section of neutron bremsstrahlung and the event rates. Following this, we will discuss the secondary particle production rates resulting from neutron-on-target interactions in Sec.~\ref{sec:neuontarget}. We will consider massive vector particles including protophobic gauge bosons at FASER/FASER2 that benefit from neutron bremsstrahlung in Sec.~\ref{sec:gaugebosons} (here, we also discuss protophobic gauge bosons produced from meson decays). Following this, in Sec.~\ref{sec:alps} we also discuss ALPs that interact with the SM photon and electron that utilize extra photons and electrons/positrons resulting from the neutron beam dump. For the ALPs scenario, we also illustrate the sensitivity of the neutron-induced ALPs at the FLArE detector (along with FASER/FASER2), which is more sensitive to low-energy signals. We finally summarize and conclude our study in Sec.~\ref{sec:conclusions}. We will show our results for an integrated luminosity of $150~\text{fb}^{-1}$ for FASER and $3~\text{ab}^{-1}$ for FASER2 and FLArE.

\medskip

\section{Neutron Bremsstrahlung} \label{sec:nbrem}
Neutron bremsstrahlung involves high-energy neutrons radiating particles while interacting (in)elastically with target nuclei. 
Bremsstrahlung contributions arising in both initial and final states are suppressed due to their destructive interference~\cite{Lebiedowicz:2013xlb} as modeled by Regge/Pomeron exchanges~\cite{Donnachie:1983ff, Donnachie:1984xq}. The most dominant bremsstrahlung effect is from initial-state radiation (ISR) when a neutron undergoes non-single diffractive (NSD) processes where the beam neutron and the target proton/neutron dissociate after scattering. 

As in the proton-initiated bremsstrahlung, the emitted particle is not only forward-directed with respect to the incoming neutron momentum but also carries most of the neutron energy. 
Thus the differential ISR cross-section $d\sigma_{\rm brem}$ can be factorized into the NSD cross-section $\sigma_{\rm NSD}$ and the differential splitting probability $dP$: 
\begin{equation}
    d\sigma_{\text{brem}} = \sigma_{\text{NSD}} \cdot dP ,
\end{equation} 
where the NSD cross-section is parametrized as $\sigma_{\text{NSD}} = 1.76 + 19.8~\big( s/\text{GeV}^2\big)^{0.057}~\text{mb}$~\cite{Likhoded:2010pc} with $s$ being the center-of-mass energy squared. 
The differential splitting probability of ISR from neutrons, under the assumption that the gauge boson is highly forward and energetic, is given by~\cite{Foroughi-Abari:2021zbm}
\begin{equation}
        dP = \omega_V(g_X, m_X, z, p_{T}^2)
        \cdot J\bigg(\frac{z, p_{T}^2}{E_X, \cos{\theta_X}}\bigg)dE_X d\cos{\theta_X},
    \label{eq:BremProb}
\end{equation}
where $E_X$, $\theta_X$, $m_X$, and $g_X$ are the energy, angle with respect to the incoming neutron, mass, and coupling strength of the gauge boson that is radiated from a neutron with momentum $p_n$ and mass $m_n$. 
We introduce $p_T$, 
the transverse momentum of the gauge boson, and $z$, the fraction of longitudinal momentum carried by the vector boson: 
\begin{equation}
    p_T = p_X\sin{\theta}, \quad 
    z = \frac{p_X\cos{\theta_X}}{p_n}
    \label{eq:z} . 
\end{equation}
In order to convert the probability into a function of $E_V$ and $\cos{\theta_V}$ (which would be suitable in order to generate the gauge boson kinematics), we include the following Jacobian in the differential probability as seen in Eq.~\eqref{eq:BremProb}:
\begin{equation}
    J\bigg(\frac{z, p_{T}^2}{E_X, \cos{\theta_X}}\bigg) = \frac{2p_X E_X}{p_n} .
    \label{eq:Jacobian}
\end{equation}
The splitting function, $\omega_V(z, p_{T}^2)$, of the neutron into a gauge boson, can be expressed as
\begin{equation}
    \begin{aligned}
        \omega_V(z, p_{T}^2) &= \frac{g_X^2}{8\pi^2} \left|F^n_{X}(m_X^2, m_n^2-H/z)\right|^2 \\ & \times  \frac{1}{H} \Bigg[z - z(1-z)  \left( \frac{2m_n^2 + m_X^2}{H}\right) + \frac{H}{2zm_X^2}\Bigg] ,
    \end{aligned}  
    \label{eq:omegaV}
\end{equation}
In the above expression, $H$ is a measure of the off-shell nature of the intermediate neutron.
\begin{equation}
    H = z[m_n^2 - (p_n - p_X)^2] .
    \label{eq:H}
\end{equation}

An important component of the splitting function in Eq.~\eqref{eq:omegaV} is the neutron form factor $F_X^n$, which is a product of the timelike form factor $F^n_{1, X}(p_X^2)$ and the hadronic form factor $F_{nn^*, X}(p^2_{n^*})$, where the former depends on the energy-momentum of the emitted particle $X$, and the latter depends on $p_{n^*}=p_n-p_X$, the momenta of the off-shell intermediate neutron. 

\begin{figure}[h]
    \centering
    \includegraphics[width=0.48\textwidth]{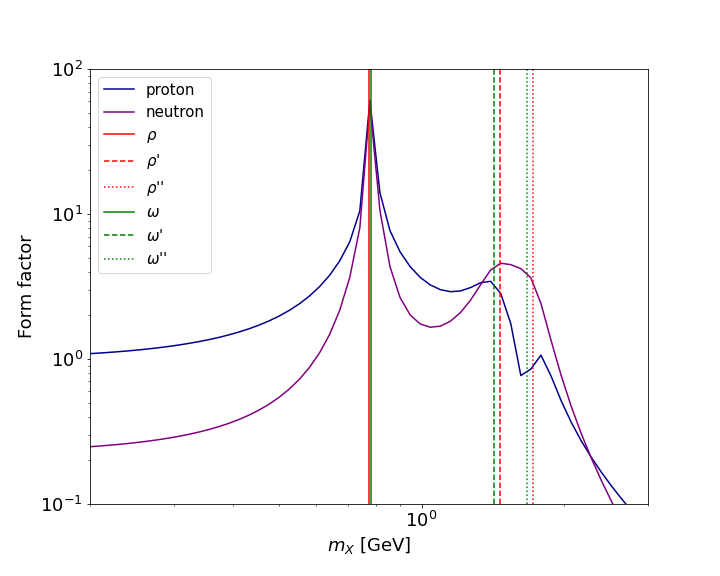}
    \captionsetup{justification=Justified}
    \caption{Hadronic form factors for protons (blue) and neutrons (purple) as a function of the mass of the emitted gauge boson $X$. The vertical lines mark the $\rho$ and $\omega$ masses where resonance is observed.}
    \label{fig:pnformfactor}
\end{figure}

The timelike form factor is obtained from the Vector Meson Dominance model~\cite{Tomasi-Gustafsson:2005svz, Faessler:2009tn, Adamuscin:2016rer} where we express the vector coupling of $X$ to the proton/neutron as a superposition of the isovector and isoscalar mesons $\rho$ and $\omega$. Using this model, the form factors for the proton and neutron are expressed as below:
\begin{widetext}
\begin{equation}
    \begin{aligned}
         F^n_{1,X}(p_X^2) &= \sum_{\omega}\frac{f_{\omega}m_{\omega}^2}{m_{\omega}^2 - p_X^2 - i m_{\omega}\Gamma_{\omega}} - \sum_{\rho}\frac{f_{\rho}m_{\rho}^2}{m_{\rho}^2 - p_X^2 - i m_{\rho}\Gamma_{\rho}}, \\
        F^p_{1,X}(p_X^2) &= \sum_{\omega}\frac{f_{\omega}m_{\omega}^2}{m_{\omega}^2 - p_X^2 - i m_{\omega}\Gamma_{\omega}} + \sum_{\rho}\frac{f_{\rho}m_{\rho}^2}{m_{\rho}^2 - p_X^2 - i m_{\rho}\Gamma_{\rho}},
        \label{eq:tformfactor} 
    \end{aligned}
\end{equation}
\end{widetext}
where the sum is over the three different $\rho$ ($\rho$, $\rho'$, $\rho''$) and $\omega$ ($\omega$, $\omega'$, $\omega''$) mesons~\cite{Workman:2022ynf}, and $f_\omega$ and $f_\rho$, commonly referred to as fit parameters, are $\{ 1.011, -0.881, 0.369\}$ and $\{ 0.616, 0.223, -0.339\}$,  respectively~\cite{Adamuscin:2016rer}. Note that the relative positive/negative sign between the $\rho$ and $\omega$ resonances in Eq.~\eqref{eq:tformfactor} gives rise to important differences in the time-like proton and neutron form factors. {Figure.~\ref{fig:pnformfactor} shows the variation of the form factors with $m_X = \sqrt{p_X^2}$. In comparison with the proton, the neutron form factor is suppressed for masses below the first $\rho$ and $\omega$ resonances ($<800~\text{MeV}$) and it is enhanced between the second and third resonances.}

The hadronic form factor $F_{nn^*, X}(p^2_{n^*})$ accounts for the off-shell nature of the intermediate neutron~\cite{Haberzettl:1998aqi, Feuster:1998cj, Oh:2000zi, Janssen:2001wk,  Penner:2002ma, Penner:2002md, Ronchen:2012eg, Kamano:2013iva}. Following Refs.~\cite{Feuster:1998cj, Foroughi-Abari:2021zbm}, we write the hadronic form factor as 
\begin{equation}
    F_{nn^*, X}(p^2_{n^*}) = \frac{\Lambda_n^4}{\Lambda_n^4 + (p^2_{n^*} - m_n^2)^2}.
\end{equation}
It does not depend strongly on the choice of the cut-off scale $\Lambda_n$, which must be larger than the mass of the neutron. We choose $\Lambda_n = 1.5~ \text{GeV}$, similar to that chosen in the proton bremsstrahlung case~\cite{Foroughi-Abari:2021zbm}.  

As mentioned earlier, the above splitting function is valid under the high-energy, highly forward assumption. These conditions can quantitatively be expressed as
\begin{equation}
        \frac{H}{4z(1-z)^2 p_p^2} \ll 1,  \quad
        p_T, m_p(m_D) \ll E_p (E_k)
    \label{eq:conditions}
\end{equation}

Where the first condition is a constraint on the off-shell nature of the intermediate neutron. The hadronic form factor manifests this condition by imposing a momentum cut-off for the off-shell neutron, where the form factor is maximized ($=1$) when the off-shell nature of the intermediate neutron (encapsulated by $p^2_{n^*} - m_n^2$) is much less than $\Lambda_n$.

After incorporating the above details into the differential cross-section of the ISR neutron bremsstrahlung, we obtain the number of radiated gauge bosons as a function of its energy ($E_X$) and angle ($\theta_X$):
\begin{equation}
    dN_{\text{brem}} = \frac{N_{A} \rho_{T} Z_{T}}{A_{T}} \lambda_{T}\  d\sigma_{\text{brem}}\,,
    \label{eq:nbrem}
\end{equation}
where $N_A$, $\rho_{T}$, $Z_{T}$, $A_{T}$, and $\lambda_{T}$ are the Avogadro's number, density, atomic number, atomic mass, and the mean free path of neutrons in target $T$ respectively. Since TAN/TAXN is composed of iron, we use $\rho_{\rm Fe} = 7.874~\text{g}/\text{cm}^3$, $A_{\rm Fe} = 56$, and $Z_{\rm Fe} = 26$. For neutron energies greater than $1~\text{GeV}$, $\lambda_{\rm Fe} \sim 11.8~\text{cm}$. Further details on simulating the radiated gauge bosons can be found in the Appendix.

 \begin{figure*}[t]
    \centering
    \begin{subfigure}[h]{.47\textwidth}
        \centering
        \includegraphics[width=\textwidth, clip]{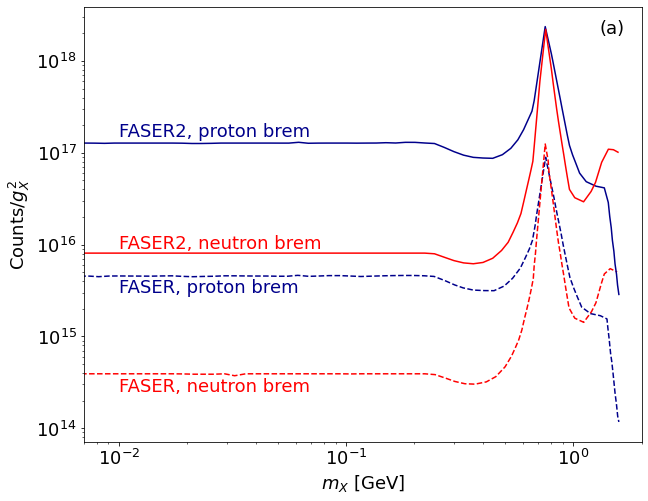}
        \captionsetup{justification=Justified, singlelinecheck=false}
        \caption{}
        \label{fig:pnbrem}
    \end{subfigure}
    \hspace{0.2cm}
    \begin{subfigure}[h]{.47\textwidth}
        \centering
        \includegraphics[width=\textwidth, clip]{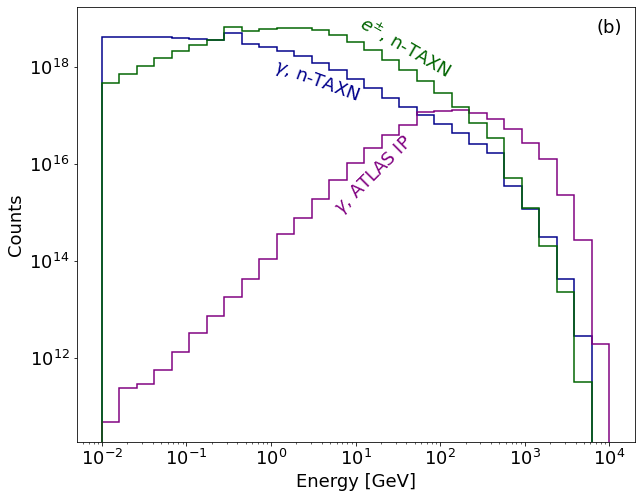}
        \captionsetup{justification=Justified, singlelinecheck=false}
        \caption{}
        \label{fig:epflux}
    \end{subfigure}\\
    \captionsetup{justification=Justified, singlelinecheck=false}
    \caption{Particle fluxes from neutron-TAN/TAXN interactions: (a) Gauge boson flux from neutron bremsstrahlung, compared with that from proton bremsstrahlung at FASER and FASER2. (b) Secondary particle fluxes (photons and $e^\pm$), compared with the photon flux produced at IP and directed towards FASER2. }
     \label{fig:neutronbrem}
\end{figure*}

The above formulation of neutron bremsstrahlung can be included in new physics searches involving new gauge bosons that have a non-trivial coupling to the neutrons such as the $U(1)_{B-L}$~\cite{FASER:2018eoc} and $U(1)_{B-3L_{i}}$~\cite{Kling:2020iar} models. This, in addition to proton bremsstrahlung, can either enhance or complement the sensitivities of various experiments

This is depicted in Figure~\ref{fig:pnbrem} where the flux of gauge bosons $X$ arising from neutron bremsstrahlung (red lines) as formulated above, is plotted in comparison to those from proton bremsstrahlung (blue lines) for the same normalized coupling. Although the center-of-mass energy of the neutron-target collision ($\sqrt{s} \sim 40~\text{GeV}$ for a $1~\text{TeV}$ neutron) is much less than that of a proton-proton collision ($\sqrt{s} = 14~\text{TeV}$), the large number of available targets within a mean free path at the dump enhances neutron bremsstrahlung for light vector gauge bosons. Along with the difference in form factor structures, this results in an $\mathcal{O}(10)$ difference in the proton versus neutron-induced flux for $m_X < 600 ~\text{MeV}$, and an enhancement for $m_X \sim 1.5~\text{GeV}$.

\medskip

\section{Particles from neutrons-on-target} \label{sec:neuontarget}
Neutrons undergoing inelastic interactions with the iron dump at TAN/TAXN produce a large flux of SM particles including photons, charged leptons, and charged and neutral mesons. 
This provides an opportunity to probe BSM physics via long-lived (boosted decay lengths greater than 20~m) or stable charged particles such as $e^{\pm}$, $\mu^{\pm}$, $\pi^{\pm}$, etc, which are unavailable at the LHC $pp$ collision because they are bent away by the magnets. We estimate the fluxes of these particles by neutrons-on-TAN/TAXN simulations done with \texttt{GEANT4}~\cite{GEANT4:2002zbu} for which we use the injection neutron energy spectrum available in \texttt{EPOSLHC}~\cite{Pierog:2013ria} and made available in the \texttt{FORESEE} package~\cite{Kling:2021fwx}.

Figure~\ref{fig:epflux} depicts the fluxes of electrons, positrons, and photons produced by neutrons that interact with TAXN and are directed towards FASER2 (similar features exist at TAN for FASER). Comparing this to the photon flux created at the IP, we observe that the neutrons induce a larger but softer flux 
at the TAN/TAXN. More quantitatively, the total number of photons induced by these interactions is $\sim10$ times larger than those from the IP. We will see that this large flux helps improve the FASER/FASER2 sensitivity of ALPs.

\medskip

\section{New gauge bosons from neutron bremsstrahlung} \label{sec:gaugebosons}
We first consider a new light vector gauge boson ($X$) of mass $m_X$, coupling to both protons and neutrons, as well as to charged leptons and neutrinos, with strength $g_X$ and charges $x_f$. In general, models involving such a gauge boson include (meant by symbol ``$\supset$'') the following operators in their Lagrangian.
\begin{equation}
    \mathcal{L}_{\rm gauge} \supset  \frac{1}{2}m_{X}^2 X_{\mu}X^{\mu} - g_{X}\sum_{f} x_f \bar{f}\gamma^{\mu}f X_{\mu} . 
    \label{eq:gblagrangian}
\end{equation}

\begin{figure}[h]
    \includegraphics[width=0.48\textwidth]{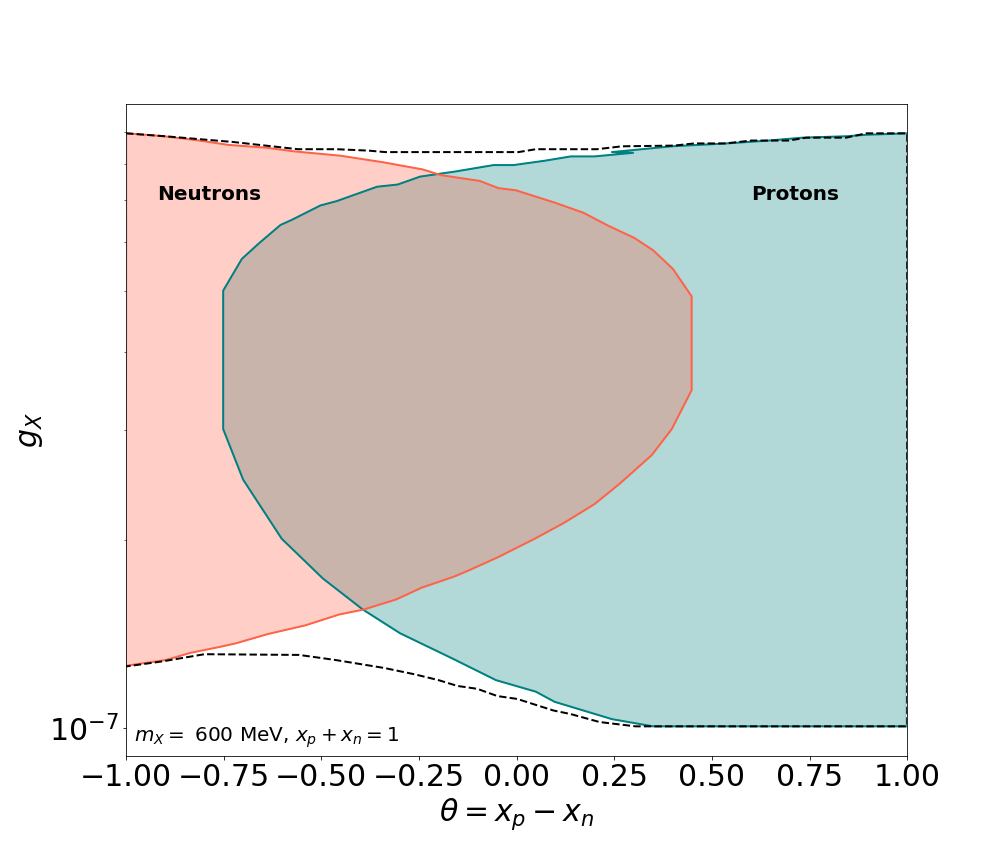}
    \captionsetup{justification=Justified, singlelinecheck=false}
    \caption{Comparison of FASER2 sensitivities with neutron and proton bremsstrahlung for different $U(1)$ charges given $m_X=600$ MeV. The dotted curve shows the total sensitivity.}
    \label{fig:pphobic_600}
\end{figure}

Figure~\ref{fig:pphobic_600} shows a 3-event contour at FASER2 for an example where the sum of charges of the proton ($x_p$) and neutron ($x_n$) are constrained to $x_p + x_n = 1$,\footnote{In terms of the fundamental Lagrangian in Eq.~\eqref{eq:gblagrangian}, this condition is $x_u + x_d = 1/3$.} $x_e = -1$, and $x_{\nu_e} = 0$. Defining a variable $\theta = x_p - x_n$, we plot the range of couplings that can be probed by proton and/or neutron bremsstrahlung for $\theta$ values from $-1$ (protophobic limit) to 1 (neutrophobic limit) for a fixed   $m_X=600~\text{MeV}$. As a result of the flux in Figure~\ref{fig:pnbrem}, the strength of couplings that can be probed through neutron bremsstrahlung is of the same order as that which can be probed through proton bremsstrahlung. The black dotted line is the combined sensitivity of proton and neutron bremsstrahlung contributions. This shows that contributions from both proton and neutron bremsstrahlung become relevant when we consider models with comparable couplings to both protons and neutrons,  such as when $\theta = 0$ (as in the $U(1)_{B-L}$ model~\cite{Marshak:1979fm, Mohapatra:1980qe, Heeck:2014zfa, Bilmis:2015lja, Jeong:2015bbi}). Therefore, at FASER2, neutron bremsstrahlung can provide additional sensitivity for gauge boson models that contain neutron couplings that are stronger or comparable to proton couplings. 

We now focus on the protophobic case, $\theta = -1$, and look at the FASER sensitivity for different vector gauge boson masses. In general, such gauge bosons can be obtained by promoting the $U(1)_B$ global symmetry to a local symmetry by $B-\gamma$ mixing. Although gauged $U(1)_B$ models are not anomaly-free while considering only SM particles, the anomaly can be canceled by introducing minimal extra fermionic matter content. This has been demonstrated in Refs.~\cite{FileviezPerez:2010gw, Dulaney:2010dj}. The coupling strength of the $X$ boson is generally written as $x_f g_X$ in terms of the baryonic and electromagnetic charges $B_f$ and $Q_f$ respectively,
\begin{equation}
    x_f = \kappa_B B_f + \kappa_{\text{em}} Q_f \, .
\end{equation}
To demonstrate the protophobic scenario, we will take $\kappa_B \sim -\kappa_{\text{em}} = 1$ (This scenario is also used to explain the Beryllium anomaly where the above condition suppressed $\pi^0$ decays as well). 
The gauge boson couplings to all fermions with a charge $x_f g_{X}$, where $x_f$ is listed in Table.~\ref{tab:pphobic}. The bounds for the choice of couplings have been explained in Ref.~\cite{Ilten:2018crw}.
\begin{table}[h]
    \begin{tabular}{| m{2cm} || m{1cm} | m{1cm} | m{1cm} | m{1.5cm} |}
        \hline 
        Fermion & $u,c,t$ & $d,s,b$  & $e$, $\mu$, $\tau$  & $\nu_e, \nu_{\mu}, \nu_{\tau}$ \\
        \hline 
        $x_f = B_f - Q_f$ & $-1/3$ & $2/3$ & $-1$ &  0  \\
        \hline 
    \end{tabular}
    \captionsetup{justification=Justified}
    \caption{Coupling values for all fermions with the protophobic gauge boson.}
    \label{tab:pphobic}
\end{table}
In this case, the charge $x_f = B_f - Q_f$, where $B_f$ is the baryon number and $Q_f$ is the electric charge. In other words, the gauge boson couples to up-type ($u$) and down-type ($d$) quarks with charges $x_{u} = -1/3$ and $x_{d} = 2/3$ respectively, while for charged leptons, $x_l = -1$~\cite{Ilten:2018crw}. 

We note that protophobic gauge bosons with $m_X\sim 17$ MeV were invoked, for instance, to explain the ATOMKI anomaly~\cite{Feng:2016jff, Feng:2016ysn, Feng:2020mbt}, but now this explanation seems to be disfavored by neutrino~\cite{Denton:2023gat} and pion decay~\cite{Hostert:2023tkg} constraints.
While neutron bremsstrahlung is an obvious source of the protophobic gauge bosons, It is important to note that lighter gauge bosons ($m_X\lesssim 500~\text{MeV}$) can appear from neutral meson decays as they have quark couplings. The branching ratio (BR) of a neutral meson decay $\mathfrak{m} \rightarrow \gamma X$ can be written as
\begin{equation}
    \text{BR}(\mathfrak{m} \rightarrow \gamma X) = 2\  \text{BR}(\mathfrak{m} \rightarrow \gamma \gamma) \beta_{\mathfrak{m}}\left(\frac{g_X}{e}\right)^2\left( 1 - \frac{m_{X}^2}{m_{\mathfrak{m}}^2}\right)^{3},
    \label{eq:BRm}
\end{equation}
where $\beta_\mathfrak{m}$ is the ratio between the photon and $X$ boson couplings of quarks in the meson  $\mathfrak{m}$. The only relevant neutral mesons in the SM with a non-negligible BR to two photons are $\pi^0, \eta$ and $\eta'$~\cite{Workman:2022ynf}.

Since the neutral pion is composed of $u$ and $d$ quarks such that $\pi^0 = (u\bar{u} - d\bar{d})/\sqrt{2}$, we observe that $\beta_{\pi^0}$ is proportional to $(Q_ux_u - Q_dx_d)^2$, where $x_q$ is given in Tab.~\ref{tab:pphobic}. We see that $(Q_ux_u - Q_dx_d)^2=0$ for the above charges. Therefore, the decay $\pi^0 \rightarrow \gamma X$ does not occur at the tree level. On the other hand, the  $\eta$ and $\eta'$ mesons are linear superpositions of the octet meson $\eta_8$ and the singlet meson $\eta_0$, with the following quark content:
\begin{equation}
    \begin{aligned}
        \eta_8  &= \frac{1}{\sqrt{6}} (u\Bar{u} + d\Bar{d} - 2s\Bar{s}), \\
        \eta_0 &= \frac{1}{\sqrt{3}} (u\Bar{u} + d\Bar{d} + s\Bar{s}).
    \end{aligned}
    \label{eq:etafunc}
\end{equation}
We then have 
\begin{equation}
    \begin{aligned}
        \eta &= \cos{\theta_p}\eta_8 - \sin{\theta_p}\eta_0 , \\
        \eta' &= \sin{\theta_p}\eta_8 + \cos{\theta_p}\eta_0 ,
    \end{aligned}
    \label{eq:etamix}
\end{equation}
where the mixing angle $\theta_p$ lies between $-17^{\circ}$ and $-20^{\circ}$~\cite{Escribano:2020jdy}. We take the average value of $\theta_p=-18.5^{\circ}$ for our numerical analysis but checked that any value within the uncertainty band essentially gives the same result. 
Using this value of $\theta_p$, the factors $\beta_{\eta}$ and $\beta_{\eta'}$ in Eq.~\eqref{eq:BRm} can be evaluated as 
\begin{eqnarray}
    \beta_{\eta} &=& \frac{\phi_{\eta}(X, \gamma)}{\phi_{\eta}(\gamma, \gamma)} = 0.29, \\
    \label{eq:etabeta}
    \beta_{\eta'} &=& \frac{\phi_{\eta'}(X, \gamma)}{\phi_{\eta'}(\gamma, \gamma)} = 1.20,
    \label{eq:etaprimebeta}
\end{eqnarray}
where the functions $\phi$ can be expressed as
\begin{eqnarray}
        \phi_{\eta}(V_1, V_2) &=& \left[ \frac{1}{f_8} \cos{\theta_p} (x^u_1 x^u_2 + x^d_1 x^d_2 - 2x^s_1 x^s_2) \right. \nonumber \\
        && \left. - \frac{\sqrt{2}}{f_0} \sin{\theta_p}(x^u_1 x^u_2 + x^d_1 x^d_2 + x^s_1 x^s_2) \right]^2,\\
    \label{eq:etaphi}
        \phi_{\eta'}(V_1, V_2) &=& \left[ \frac{1}{f_8} \sin{\theta_p} (x^u_1 x^u_2 + x^d_1 x^d_2 - 2x^s_1 x^s_2) \right. \nonumber \\
        && \left. + \frac{\sqrt{2}}{f_0} \cos{\theta_p}(x^u_1 x^u_2 + x^d_1 x^d_2 + x^s_1 x^s_2) \right]^2.
    \label{eq:etaprimephi}
\end{eqnarray}
Here $f_0=1.05f_\pi$ and $f_8=1.3f_\pi$~\cite{Nefkens:2002sa}, with $f_\pi$ being the pion decay constant~\cite{Workman:2022ynf}. Also, $ x^q_i$ is the charge carried by quark $q = u, d, s$ with the gauge boson $V_i$ (where $i = 1, 2$ for $X$ and $\gamma$), i.e.~$x^q_X\equiv x_qQ_q$ and $x^q_{\gamma} \equiv Q_q$. 

\begin{figure}[h]
    \includegraphics[width=0.48\textwidth]{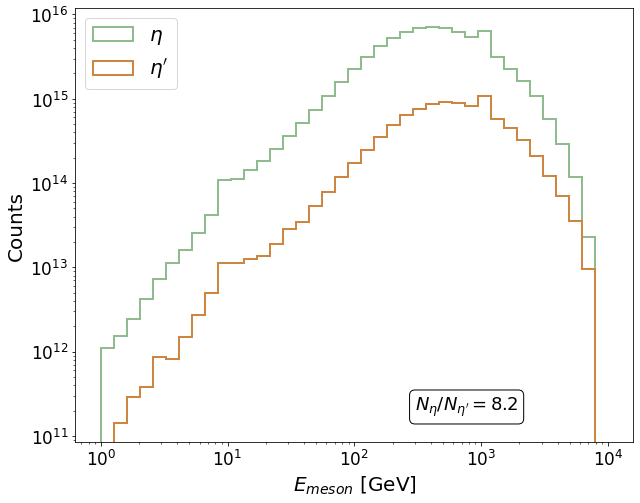}
    \captionsetup{justification=Justified}
    \caption{Energy flux of $\eta$ and $\eta'$ mesons that are towards FASER2 detector normalized to a total luminosity of $3~\text{ab}^{-1}$.}
    \label{fig:etafluxes}
\end{figure}

Therefore, the branching ratios of $\eta/\eta' \rightarrow X \gamma$ are given by 
\begin{equation}
    \begin{aligned}
        \text{BR}(\eta \rightarrow X \gamma) &= 2\times0.39\times0.29 \bigg(\frac{g_X}{e} \bigg)^2 \bigg( 1 - \frac{m_{X}^2}{m_{\eta}^2}\bigg)^{3}, \\
        \text{BR}(\eta' \rightarrow X \gamma) &= 2\times0.02\times1.20 \bigg(\frac{g_X}{e} \bigg)^2 \bigg( 1 - \frac{m_{X}^2}{m_{\eta'}^2}\bigg)^{3} , 
    \end{aligned}
    \label{eq:etabr}
\end{equation}
using the fact that $\text{BR}(\eta \to \gamma \gamma)=0.39$ and $\text{BR}(\eta' \to \gamma \gamma)=0.02$ in the SM~\cite{Workman:2022ynf}. This implies that the $\eta$ decay contribution to the protophobic gauge boson production is about 5 times stronger than that from the $\eta'$ decay. This, together with the fact that the $\eta$ production rate is $\sim 8$ times higher than that of $\eta'$ (as seen in Figure~\ref{fig:etafluxes}), implies that it is okay to drop the $\eta'$ contribution from our sensitivity analysis. We use the meson fluxes 
generated by \texttt{EPOSLHC}~\cite{Pierog:2013ria} and made available in the \texttt{FORESEE} package~\cite{Kling:2021fwx}. To check the uncertainty on the fluxes, we compared the fluxes of $\eta$, $\eta'$, and neutrons generated by \texttt{EPOSLHC} and  \texttt{QSCJET}, but found them to be very similar.

\begin{figure}[h]
    \includegraphics[width=0.48\textwidth]{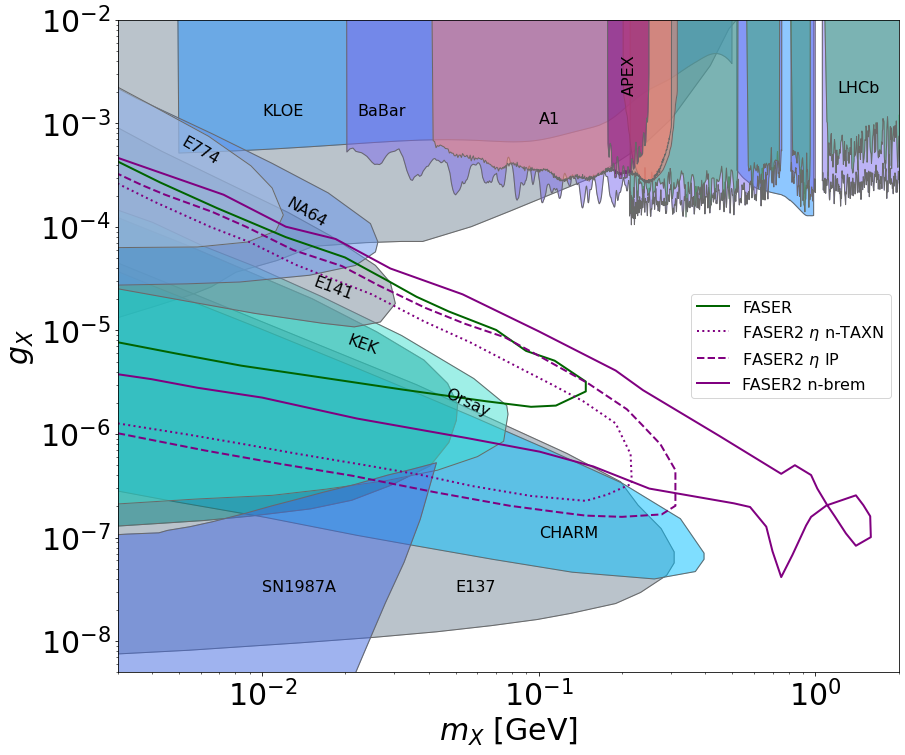}
    \captionsetup{justification=Justified, singlelinecheck=false}
    \caption{FASER/FASER2 sensitivities for protophobic gauge bosons. FASER2 curves are broken down for different sources. Shaded regions show existing bounds.}
    \label{fig:pphobic_sens}
\end{figure}

Figure~\ref{fig:pphobic_sens} shows the 3-event contour (95\% CL sensitivity) in the gauge boson mass-coupling plane at FASER (green) and FASER2 (purple). We find that the high-energy $\eta$ and neutrons at FASER facilitate searches for protophobic gauge bosons in a region that has not been previously explored by beam-dump and other experiments (shaded regions, taken from Ref.~\cite{Ilten:2018crw}). The sensitivity will be further improved at FASER2 with higher luminosity. The different contributions to FASER2 are broken down to show that neutron bremsstrahlung (solid) plays a vital role in probing the 
middle-right region of parameter space (short lifetimes) for all masses. 
The reason is that neutron-induced gauge bosons are produced at TAN/TAXN, not at IP, i.e., closer to the detector, and are more energetic. The ``beak'' shape at the high-mass edge corresponds to the peak in the bremsstrahlung flux in Figure~\ref{fig:neutronbrem}a at $m_X \simeq 780$ MeV, which is due to the $\rho$ meson resonance. {The other ``head'' feature that appears beyond $1.1$~GeV is a result of the neutron form factor which peaks between the second and third $\rho$ and $\omega$ resonances as seen in Figure~\ref{fig:pnformfactor}. We hence realize that baryophilic gauge bosons with masses around 1.5~GeV can be probed only via neutron bremsstrahlung.}

We also include the protophobic gauge bosons that appear from the decay of $\eta$ meson decays as shown by the dashed and dotted purple lines, corresponding to the $\eta$ mesons produced at the IP (from $pp$ collisions) and at TAXN (from neutron interactions with the dump), respectively. Although gauge bosons appearing from these decays are the most dominant for $m_X<m_\eta=540~\text{MeV}$, they only probe parameters that have already been excluded by electron beam-dump experiments. Therefore, we realize that protophobic gauge bosons from neutron bremsstrahlung give us the most insight. 

For comparison, we also show the existing experimental constraints on the protophobic gauge boson model by the shaded regions, taken from Ref.~\cite{Ilten:2018crw}, and supernova limits from Ref.~\cite{Rrapaj:2015wgs}. The grey-shaded bounds arise from $\Upsilon$~\cite{Aranda:1998fr, ARGUS:1986nzm} and $\eta$ decays~\cite{Tulin:2014tya, Prakhov:2008zz}, from longitudinal enhancements~\cite{Dror:2017nsg, Dror:2017ehi}, in $B_{u,d} \rightarrow K X$ decays ~\cite{Belle:2017oht}, $K \rightarrow \pi X$~\cite{KTeV:2003sls, E949:2008btt}, $Z \rightarrow X \gamma$~\cite{L3:1997exg, DELPHI:1996qcc}, and the observational lack of new anomaly-canceling fermions~\cite{Cline:2014dwa, Dobrescu:2014fca, Dror:2017nsg}. The electron beam-dump constraints come from KEK~\cite{Konaka:1986cb}, E137~\cite{Bjorken:1988as}, E141~\cite{Riordan:1987aw}, E774~\cite{Bross:1989mp}, NA64~\cite{NA64:2018lsq}, and Orsay~\cite{Davier:1989wz}), while the  proton beam-dump constraints are from CHARM~\cite{CHARM:1985anb}. The $e^+ e^-$ collider constraints from BaBar~\cite{BaBar:2014zli} and KLOE~\cite{KLOE-2:2016ydq}, $pp$ collider constraint from LHCb~\cite{LHCb:2017trq} and others such as A1~\cite{Merkel:2014avp} and APEX~\cite{APEX:2011dww} are also depicted in the sensitivity plot.

Although Figure~\ref{fig:pphobic_sens} is obtained without imposing energy thresholds on the final state leptons and hadrons, we observe that the contour remains unaffected for energy thresholds that are as high as 80-100 GeV~\cite{Berlin:2018jbm, FASER:2018eoc}. This is because the majority of the gauge bosons have energies between $\mathcal{O}(0.1 - 10)$~TeV.

\medskip

\section{ALPs with neutron dump} \label{sec:alps} 
ALPs with couplings to photons have been studied at FASER~\cite{Kling:2022ehv, Feng:2018pew}, where high-energy photons produced at the IP convert to ALPs through the Primakoff scattering at the TAN/TAXN. These ALPs, which are very energetic with momentum $\mathcal{O}(1~\text{TeV})$, allow FASER to explore new parameter space toward the middle-right region beyond the existing beam-dump constraints.

While the FASER sensitivity to ALPs interacting with SM fermions has been studied using heavy meson decays and Yukawa-like coupling structure~\cite{Feng:2022inv}, ALPs with electron couplings have received less attention. We emphasize that there are plenty of electrons and positrons created at the TAN/TAXN due to neutron interactions. As we show here, these can be utilized to study ALPs with electron couplings at FASER. In addition, low-energy photons produced at the TAN/TAXN facilitate Compton-like scattering processes, hence serving as a factory of electrophilic ALPs. With $a$ denoting the ALP field, the relevant Lagrangian for the ALPs is 
\begin{equation}
    \mathcal{L}_{ae} \supset -\frac{1}{2}m_a^2 a^2 -ig_{ae}a\bar{e}\gamma^5 e,
    \label{eq:aelag}
\end{equation}
from which we identify three major production channels:
i) Compton-like scattering ($\gamma e^- \rightarrow a e^-$)~\cite{AristizabalSierra:2020rom}, ii) associated production ($e^+ e^- \rightarrow a \gamma$), and iii) electron-positron annihilation ($e^+ e^- \rightarrow a$)~\cite{Nardi:2018cxi, CCM:2021jmk}.  
These ALPs are primarily detected when they decay to electron-positron pairs. Kinematically, only those ALPs heavier than $2m_e\simeq 1$ MeV can decay to $e^+e^-$ pair.

\begin{figure}[t!]
    \centering
    \includegraphics[width=0.48\textwidth]{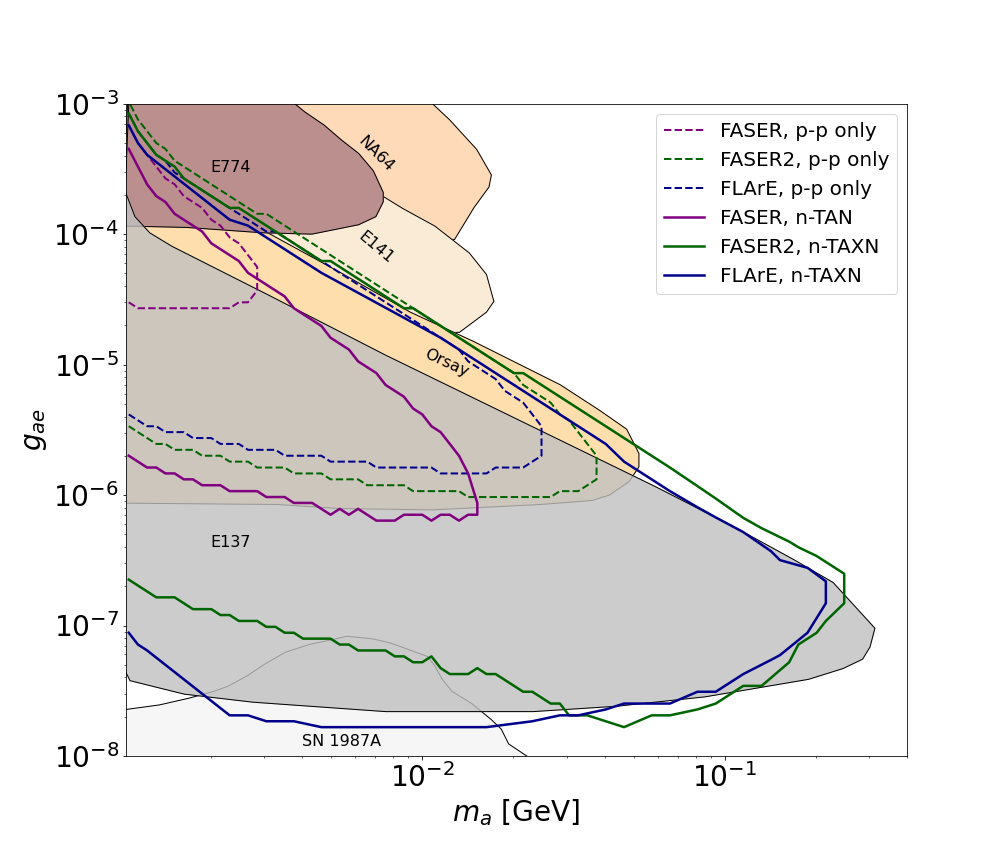}
    \captionsetup{justification=Justified}
    \caption{New FASER/FASER2 and FLArE sensitivity to ALP-electron coupling. Shaded regions show existing bounds.
    }
    \label{fig:alpesens}
\end{figure}

Figure~\ref{fig:alpesens} exhibits our sensitivity estimates for these ALPs that appear from secondary photons, electrons, and positrons induced by neutron interactions. In this scenario, we also demonstrate the sensitivity at the distant FLArE detector (620~m away from the ATLAS IP). The green (purple) lines are for FASER (FASER2) and the dark blue lines are for FLArE. The dashed lines are sensitivities only with ALPs produced by IP-origin photons, whereas the solid lines are those produced from neutron-induced secondary particles. While we depict the sensitivities by applying a 1~GeV energy threshold for the FASER detectors and a 30~MeV~\cite{Kling:2022ykt}  threshold for FLArE, we notice that the ceiling of the sensitivity plot remains unaffected for larger energy thresholds (up to 100~GeV for FASER and 100-MeV for FLArE). We find that the latter sensitivities are significantly better due to the low-energy photons as well as the additional flux of electrons and positrons. The existing constraints on the ALP parameter space are shown by the shaded regions, where the dominant limits come from previous beam-dump experiments, collected in Ref.~\cite{AxionLimits}. These include E137~\cite{Bjorken:1988as, Andreas:2010ms}, Orsay~\cite{Bechis:1979kp}, E141~\cite{Riordan:1987aw}, E774~\cite{Bross:1989mp}, and NA64~\cite{Andreev:2021fzd}. The supernova bound was taken from Ref.~\cite{Lucente:2021hbp}, which however depends on other assumptions and can in principle be evaded~\cite{Khoury:2003aq, Jain:2005nh, Masso:2006gc,  Mohapatra:2006pv, Dupays:2006dp, Brax:2007ak, Jaeckel:2006xm, DeRocco:2020xdt, Sung:2021swd}.

Thus we see that neutron-on-target interactions have a large impact on ALPs with electron couplings. We also find that FASER2 and FLArE complement each other. FASER2, being the larger and closer detector, can probe the higher mass and coupling region ($m_a\sim 100~\text{MeV},~g_{ae} \sim 10^{-6}$), whereas FLArE, due to its lower energy threshold, can probe lower masses and weaker couplings ($m_a\sim 30~\text{MeV},~g_{ae} \sim 2\times 10^{-8}$).

\medskip

\section{Conclusions} \label{sec:conclusions}
This work demonstrates with two examples how neutrons can be used for the first time for probing new physics at a high-energy dump facility like FASER. 
Copious amounts of high-energy neutrons from ATLAS IP propagate past the LHC magnets and radiate new particles on interactions with iron dump.

We derived the FASER2 sensitivity for protophobic gauge bosons, which extends up to $\sim 1$ GeV in the gauge boson mass when neutron bremsstrahlung is considered. 
Beyond the photophobic scenario, neutrons can provide an additional handle for any model that involves new particles with neutron couplings.   

Furthermore, since neutron-on-target interactions are the only source of long-lived charged particles at FASER, they open up additional possibilities to probe BSM particles coupling to these secondary particles. 
 Utilizing secondary $e^{\pm}$ and photons, we took ALPs coupling to electrons as an example to show that FASER2 can probe new parameter space in the range of $g_{ae}\sim 10^{-8}-10^{-6}$, and $m_a\sim 20-300~\text{MeV}$. We reserve the impact of secondary productions on various other light mediators, such as ALPs with photon couplings, leptophilic scalars, etc. for future work.

\medskip

\section*{Acknowledgments}
We thank Brian Batell, Joshua Berger, and Felix Kling for useful discussions. The work of PSBD is supported in part by the U.S.~Department of Energy under Grant No.~DE-SC0017987.
The work of BD, AK, DK, and HK is supported in part by the U.S.~Department of Energy Grant DE-SC0010813. 
The work of TH is supported in part by the U.S.~Department of Energy under grant No.~DE-SC0007914 and by the PITT PACC.

\medskip
\appendix

\section{Simulation Scheme}

To simulate the gauge bosons from two-body meson decays for a particular gauge boson mass, we utilize the branching ratio in Eq.~\eqref{eq:etabr} as a weight factor for each gauge boson. To simulate event kinematics, we first work in the rest frame of the decaying $\eta$ meson where the final state energies are fixed, and the angular distribution is uniform. We construct the four-momentum of $X$ in the rest frame by randomly choosing $\cos{\theta}$ and $\phi$, where $\theta$ and $\phi$ are the polar and azimuthal angles respectively, and then boost it to the laboratory frame to generate events. 

Simulating gauge bosons from neutron bremsstrahlung is more involved. Although Eq.~\eqref{eq:nbrem} gives the number of bremsstrahlung products, all the information about energies and angles of the gauge boson is encapsulated in Eq.~\eqref{eq:BremProb}. Therefore, we simulate gauge bosons from the latter equation and then normalize it to the former one. Since this probability is under the assumption that the gauge bosons are highly forward-going and energetic, the probability function is highly peaked at $\cos{\theta_X} = 1$. In order to numerically integrate over $\cos{\theta_X}$, we pick only the peak value and integrate $\cos{\theta_X}$ over the angular acceptance of the detector, i.e., if $\theta_n$ is the polar angle of the neutron and $\theta_d$ is the angle subtended by the detector at the point where the neutron radiates $X$, the limits of the integral are $\cos{(\theta_d - \theta_n)} < \cos{\theta_X} < 1$. For the energy integral, the distribution is not highly-peaked at a specific value. Rather, we find that to a good approximation it linearly increases with $E_X$ for high energies. To select the energies of these radiated bosons for a given neutron energy $E_n$, we pick $ E_X$ from a linear distribution ranging from $E_n/2$ to $E_n$ based on the above assumptions.

\bibliography{references}

\end{document}